\documentclass[10pt]{article}

\usepackage{amsfonts,amssymb,amsmath}
\usepackage{xspace}

\newcommand{\NAND}{\mbox{\sc nand}\xspace}
\newcommand{\be}{\begin{equation}}
\newcommand{\ee}{\end{equation}}
\newcommand{\ket}[1]{|#1\rangle}
\newcommand{\norm}[1]{\|#1\|}

\begin{document}

\title{\LARGE\bf Discrete-query quantum algorithm for {\Large\bf NAND} trees}

\author{
Andrew M.~Childs%
\thanks{Institite for Quantum Information, California Institute of Technology}
\and Richard Cleve%
\thanks{David R.~Cheriton School of Computer Science and Institute
for Quantum Computing, University of Waterloo}
\thanks{Perimeter Institute for Theoretical Physics}
\and
Stephen P.~Jordan%
\thanks{Center for Theoretical Physics, Massachusetts Institute of Technology}
\and David Yonge-Mallo$^{\dag}$}

\date{16 February 2007}

\maketitle \thispagestyle{empty}

\vspace{-2ex}
\begin{abstract}
\noindent Recently, Farhi, Goldstone, and Gutmann gave a quantum
algorithm for evaluating \NAND trees that runs in time $O(\sqrt{N
\log N})$ in the Hamiltonian query model.  In this note, we point
out that their algorithm can be converted into an algorithm using
$O(N^{1/2 + \epsilon})$ queries in the conventional quantum query
model, for any fixed $\epsilon > 0$.
\end{abstract}

\bigskip

\noindent A \NAND tree of depth $n$ is a balanced binary tree whose
internal vertices represent \NAND gates.  Placing bits
$x_1,\ldots,x_{2^n}$ at the leaves, the root of the \NAND tree
evaluates to the function $f_n(x_1,\ldots,x_{2^n})$, where $f_n :
\{0,1\}^{2^n} \to \{0,1\}$ is defined recursively as follows.  For
$n=0$, $f_0(x) = x$, and for $n > 0$, \be
  f_n(x_1,\dots,x_{2^n})
  = \neg\big(f_{n-1}(x_1,\dots,x_{2^{n-1}})
  \wedge f_{n-1}(x_{2^{n-1}+1},\dots,x_{2^{n}})\big).
\ee The goal of the \NAND tree problem is to evaluate
$f_n(x_1,\dots,x_{2^n})$, making as few queries to the bits
$x_1,\dots,x_{2^n}$ as possible.
The optimal classical randomized algorithm for this problem makes
$\Theta(N^{0.753})$ queries, where $N = 2^n$ \cite{SW86,Santha91,Snir85}.  Until now, no
better quantum algorithm was known, whereas the best known quantum
lower bound is only $\Omega(\sqrt N)$ \cite{BS04}.  Here we  show
that for any fixed $\epsilon>0$, the quantum query complexity of
evaluating $\NAND$ trees is $O(N^{1/2+\epsilon})$.

Very recently, Farhi, Goldstone, and Gutmann \cite{FGG07} proposed a
quantum algorithm that evaluates $\NAND$ trees in time $O(\sqrt{N
\log N})$, albeit in the unconventional Hamiltonian oracle
model~\cite{FG98,Mochon06}.  In their version of this model, we are given
access to a Hamiltonian $H_O$ acting on $n+1$ qubits as \be
  H_O \ket{b,k} = -x_k \ket{\neg b,k}
\ee for all $b \in \{0,1\}$ and $k \in \{0,1\}^n$, and the goal is
to perform the computation using evolution according to $H_O +
H_D(t)$ for as short a time as possible, where $H_D(t)$ is an
arbitrary driving Hamiltonian (that is possibly time-dependent and may
act on an extended Hilbert space).

In the conventional quantum query model, the input is accessible via
unitary operations of the form \be\label{eq:query-oracle}
  U_O \ket{k,a} = \ket{k,a \oplus x_k}.
\ee Two queries of $U_O$ can be used to implement evolution
according to $H_O$ for an arbitrary time $t$, which can be seen as
follows.
The procedure acts on states of the form $\ket{b,k,a}$ (where the
last register is an ancilla qubit), and consists of the following
steps.  First, apply $U_O$ to the second and third registers.  Then
apply a controlled-$R(t)$ gate with the first register as the target
and the third register as the control, where \be R(t) = \left(
\begin{array}{lr}
\cos t & i\sin t \\
i\sin t & \cos t
\end{array}
\right). \ee Finally, apply $U_O$ to the second and third registers
again. With the ancilla qubit initially in the $\ket{0}$ state, the net
effect of this procedure is the mapping
$\ket{b,k,0} \mapsto \cos(x_k t)\ket{b,k,0} + i \sin(x_k t)\ket{\neg b,k,0}$,
which corresponds to evolution by $H_O$ for time $t$ (that is, the
unitary operation $e^{-iH_O t}$).

This simulation of $H_O$ does not imply that any fast
algorithm in the Hamiltonian oracle model can be turned into an
algorithm with small query complexity in the conventional quantum
query model.  Accurate simulation of the evolution according to $H_O
+ H_D(t)$ apparently requires many interleaved evolutions of $H_O$
and $H_D(t)$ each for a small time, yet each of which requires two
unitary queries to simulate.  Nevertheless, it turns out that a
Hamiltonian of the kind used in \cite{FGG07} {\em can} be simulated
in the conventional quantum query model with only small overhead.

In the algorithm of \cite{FGG07}, 
$H_D(t)$
is time-independent, so the evolution for time $t$ is given
by $e^{-i(H_O + H_D)t}$.  Such evolution according to a sum of
time-independent Hamiltonians can be simulated using a high-order
approximation of the exponential of a sum in terms of a product of
exponentials of the individual terms.
As noted in~\cite{BACC05,Chi04}, by using a $p^{\mbox{\scriptsize th}}$
order approximation, the simulation can be performed in
$O((ht)^{1+1/2p})$ steps, where $h = \max\{\norm{H_O},\norm{H_D}\}$,
and where the constant depends on $p$ and on the desired simulation
error.  We have $\norm{H_O} \le 1$ and
$\norm{H_D} \le 3$, so $h \le 3$.  Choosing $p$ to be an arbitrarily large
constant, we obtain a simulation using $O(t^{1+\delta})$ steps,
where $\delta=1/2p$ is arbitrarily small.  Since the algorithm of
\cite{FGG07} applies $H$ for time $t=O(\sqrt{N \log N})$, it follows
that the corresponding time evolution can be simulated using
$O(N^{1/2+\epsilon})$ queries to the original oracle, where $\epsilon>0$ may be arbitrarily small.

The result can also be deduced by noting that, given query access to
the inputs via $U_O$ (Eq.~\ref{eq:query-oracle}), one can easily
simulate an oracle for the matrix elements of the underlying Hamiltonian $H_O + H_D$ used in~\cite{FGG07}, and then apply results in~\cite{BACC05,Chi04} for
simulating sparse Hamiltonians.

\section*{Acknowledgments}

AMC received support from the U.S.\ NSF under grant no.\
PHY-0456720, and from the U.S.\ ARO under grant no.\
W911NF-05-1-0294. RC received support from Canada's CIAR, MITACS,
NSERC, and the U.S.\ ARO/DTO. SPJ received support from the U.S.\
ARO/DTO's QuaCGR program. DY received support from the U.S.\
ARO/DTO.


\end{document}